\pdfoutput=1
\documentclass[11pt, oneside]{article}   
\usepackage{geometry}
\usepackage{graphicx}
\usepackage{color}
\usepackage{xcolor}
\usepackage{amssymb}
\usepackage{hyperref}
\usepackage{wrapfig}
\usepackage{caption}
\usepackage{booktabs}
\usepackage{amsmath,amsfonts,bm, bbm}
\usepackage[edges]{forest}

\definecolor{foldercolor}{RGB}{200,200,200}

\tikzset{pics/folder/.style={code={%
    \node[inner sep=0pt, minimum size=#1](-foldericon){};
    \node[folder style, inner sep=0pt, minimum width=0.3*#1, minimum height=0.6*#1, above right, xshift=0.05*#1] at (-foldericon.west){};
    \node[folder style, inner sep=0pt, minimum size=#1] at (-foldericon.center){};}
    },
    pics/folder/.default={20pt},
    folder style/.style={draw=foldercolor!80!black,top color=foldercolor!40,bottom color=foldercolor}
}

\forestset{is file/.style={edge path'/.expanded={%
        ([xshift=\forestregister{folder indent}]!u.parent anchor) |- (.child anchor)},
        inner sep=1pt},
    this folder size/.style={edge path'/.expanded={%
        ([xshift=\forestregister{folder indent}]!u.parent anchor) |- (.child anchor) pic[solid]{folder=#1}}, inner xsep=0.75*#1},
    folder tree indent/.style={before computing xy={l=#1}},
    folder icons/.style={folder, this folder size=#1, folder tree indent=3*#1},
    folder icons/.default={12pt},
}

\title{Instructions and Guide for \\ Diagnostic Questions:
The NeurIPS 2020 Education Challenge}
\author{
Zichao Wang $^1$ \thanks{
Equal contribution, co-first authors
$\dagger$ co-senior authors
$^1$Rice University,
$^2$University of Cambridge,
$^3$Eedi,
$^4$Microsoft Research,
$^5$Alan Turing Institute.
}, Angus Lamb$^{4 *}$, Evgeny Saveliev$^4$,\\ Pashmina Cameron$^4$, Yordan Zaykov$^4$, Jos\'{e} Miguel Hern\'{a}ndez-Lobato$^{245}$,\\ Richard E. Turner$^{24}$,  Richard G. Baraniuk$^1$, Craig Barton$^3$,  Simon Peyton Jones$^4$,\\Simon Woodhead$^{3\dagger}$, Cheng Zhang$^{4\dagger}$}

\date{}

\begin{document}
\maketitle

\begin{abstract}
Digital technologies are becoming increasingly prevalent in education, enabling personalized, high quality education resources to be accessible by students across the world. 
Importantly, among these resources are \textit{diagnostic questions}: the answers that the students give to these questions reveal key information about the specific nature of misconceptions that the students may hold. 
Analyzing the massive quantities of data stemming from students' interactions with these diagnostic questions can help us more accurately understand the students' learning status and thus allow us to automate learning curriculum recommendations. In this competition, participants will focus on the students' answer records to these multiple-choice diagnostic questions, with the aim of 1) accurately predicting which answers the students provide; 2) accurately predicting which questions have high quality; and 3) determining a personalized sequence of questions for each student that best predicts the student's answers.
These tasks closely mimic the goals of a real-world educational platform and are highly representative of the educational challenges faced today. We provide over 20 million examples of students' answers to mathematics questions from Eedi, a leading educational platform which thousands of students interact with daily around the globe. 
Participants to this competition have a chance to make a lasting, real-world impact on the quality of personalized education for millions of students across the world.
\end{abstract}

\subsection*{Keywords}
Personalized Education, Unsupervised Learning, Missing value prediction, Active Learning

\tableofcontents

\section{Introduction}

\subsection{Background and impact}

\paragraph{Background}

The prevalence of free or affordable online education systems is making high quality education available to a wider audience. On these platforms students can learn by watching instructional videos, reading (possibly interactive) course materials, and talking with other students and mentors in learning forums. To measure student understanding many of these platforms include an assessment component. By mining the data collected by these assessments, we can, in theory, extract useful educational information such as how students are learning, and recommend suitable learning interventions to improve learning outcomes. However, the quality of the insights derived is dependent on the quality of the questions in the assessments. 

\textit{Formative assessment} is concerned with the careful design of assessments which elicit detailed information that can be used to improve instruction and student learning while it is happening. A deceptively simply but powerful question type used for formative assessment is a \textit{diagnostic question}.

A diagnostic question is a multiple-choice question with four answers, exactly one of which is correct and where each of the three incorrect answers is chosen to highlight a common misconception. If a student gets a diagnostic question wrong, educators are not left to guess why. The student's choice of incorrect answer reveals something about the nature of their misconception which is valuable information in seeking to help them resolve it \cite{wylie2006dqs}. Diagnostic questions can be constructed to induce retrieval of information pertaining to the incorrect alternatives. Therefore, students are not only challenged to consider why the right answer is right, but why the wrong answers are wrong \cite{little2018retrieval}. 

It is challenging to write good diagnostic questions, where each of the incorrect answers is a plausible distractor. Even experienced teachers find the task time-consuming. To address this challenge, a platform was created (\url{https://diagnosticquestions.com/}) by the team behind Eedi (\url{https://eedi.com}), to facilitate teachers crowd-sourcing diagnostic questions. Inevitably, there is variation in the quality of the questions created. 

When teachers create diagnostic questions, each incorrect answer should be chosen to highlight a common misconception, but these misconceptions are not labelled or linked between questions. It is entirely possible for an incorrect answer to be chosen so poorly that it is obviously wrong, and therefore it will never be chosen by a student.

In order to diagnose student learning accurately it is essential that they are presented with good questions. A good diagnostic question identifies the specific nature of a student’s misconception. They need to be unambiguous, and crucially students should not be able to get them correct whilst still holding a key misconception. Moreover, teaching and learning time is limited, so we need to prioritise those questions which capture the most information about the student's knowledge and misconceptions. 

\paragraph{Summary of competition and impact on education} 
In this competition, we challenge participants to develop novel methodologies to understand and improve students’ learning and measure the quality of diagnostic questions. There are four tasks, described in Section~\ref{sec:tasks} and summarized here: 

\begin{enumerate}

\item The first task is to predict whether or not students will answer questions correctly. \newline {\bf Real-world impact:} Enable recommending questions of an appropriate difficulty to a given student that best fit their background and learning status.

\item The second task extends this to the prediction of which answer students choose for each question.  \newline {\bf Real-world impact:} Enable discovering potential common misconceptions that students have by clustering of question-answer pairs which may indicate the same or related misconceptions.

\item The third task is to devise a metric to measure the quality of the questions. This metric will be evaluated against the opinions of domain experts. \newline {\bf Real-world impact:} Enable  feedback to be provided to authors of diagnostic questions so they can revise poor quality questions and to guide teachers to choose questions for their students.

\item The fourth task is to acquire a limited set of answers from students  for student performance prediction on unseen questions. This requires personalized machine learning methods with estimation of value of information. \newline {\bf Real-world impact:} Enable personalized assessments for each student to improve learning outcomes.

\end{enumerate}

The competition provides an in-depth introduction to educational data mining because our tasks mimic the learning analytics and personalization tasks common to many education platforms. The competition will use data from an educational platform that is already deployed and used at scale.  The data describes real answers given by real students to real questions.  By providing the opportunity to work on genuine educational data and real problems in an engaging manner, our competition will attract talent to the important field of machine learning in education.

We expect the competition to bring fundamental advances to educational data mining technologies, particularly those that analyze students' learning progress and recommend personalized learning curricula. These methods will be deployed in a real educational platform where they will improve the learning outcomes of millions of students.

\paragraph{Impact on the machine learning community}

The competition involves multiple fundamental machine learning challenges that need to be addressed. Some challenges are common in recommender systems but appear in the context of educational data mining: how to deal with the sparsity of the data because each student answered only a small fraction of all questions? How to effectively use student and question metadata, such as student demographics, to improve the prediction? Other challenges are reminiscent of active learning: how to optimally select the sequence of questions in order to maximize prediction accuracy? Another challenge is how to effectively perform matrix completion for unordered, categorical data. Tackling these challenges in the unique context of educational data mining will be of significant technical interest to the NeurIPS community and, more broadly, the machine learning community as a whole.

\subsection{Navigating the Competition}
This document serves both as an introduction to the competition, and as a guide for participants to navigate the competition. {\bf The remainder of this document contains important information on various aspect of the competition}. We encourage participants to become familiar with, and regularly refer back to, the following information during the competition: 
\begin{itemize}
    \item A master list of files that participants will receive and where to download them (Section~\ref{sec:list});
    \item Detailed descriptions of the data files (Section~\ref{sec:data});
    \item Detailed descriptions of each task in the competition including evaluation metric(s) for each task (Section~\ref{sec:tasks});
    \item Instructions for submitting each task to CodaLab (Section~\ref{sec:submission});
    \item A getting started guide providing scripts to load the data, perform local evaluation and prepare sample submissions (Section~\ref{sec:get-started}).
\end{itemize}

\subsection{Competition File List}
\label{sec:list}
Below we provide a list of the files available to participants. The files below are in two zip files, both of which can be found in the CodaLab competition website. Click on the \texttt{Participate} tab, then \texttt{Get Data}.
\paragraph{Public Data} Contains the data folder which contains three sub-folders (details of each file in Section~\ref{sec:data}):

\begin{forest}
  for tree={font=\ttfamily, grow'=0,
  folder indent=1.5em, folder icons}
  [data
    [images
      [{0.jpg, 1.jpg, ..., 947.jpg}, is file]]
    [metadata
      [answer\_metadata\_task\_1\_2.csv, is file]
      [answer\_metadata\_task\_3\_4.csv, is file]
      [question\_metadata\_task\_1\_2.csv, is file]
      [question\_metadata\_task\_3\_4.csv, is file]
      [student\_metadata\_task\_1\_2.csv, is file]
      [student\_metadata\_task\_3\_4.csv, is file]
      [subject\_metadata.csv, is file]
      ]
    [train\_data
      [train\_task\_1\_2.csv, is file]
      [train\_task\_3\_4.csv, is file]]
  ]
\end{forest}

\paragraph{Starting Kit} Contains the folder \texttt{starter\_kit} to help participants get started on data loading, local evaluation and submission preparation:

\begin{forest}
  for tree={font=\ttfamily, grow'=0,
  folder indent=1.5em, folder icons}
  [starter\_kit
    [submission\_templates
        [submission\_task\_1\_2.csv, is file]
        [submission\_task\_3.csv, is file]
        [submission\_model\_task\_4.py, is file]
    ]
    [task\_1
      [local\_data\_split.py, is file]
      [local\_evaluation.py, is file]
      [sample\_model\_majority.py, is file]]
    [task\_2
      [local\_data\_split.py, is file]
      [local\_evaluation.py, is file]
      [sample\_model\_majority.py, is file]]
    [task\_3
      [local\_evaluation.py, is file]
      [sample\_model\_entropy.py, is file]]
    [task\_4
      [local\_data\_split.py, is file]
      [numpy
        [local\_evaluation.py, is file]
        [model.py, is file]
        [train\_model.py, is file]]
      [pytorch
        [local\_evaluation.py, is file]
        [model.py, is file]
        [train\_model.py, is file]]]
  ]
\end{forest}

\section{Data}
\label{sec:data}

\begin{table}[h]
\vspace{15.0pt}
\centering
\captionsetup{width=0.9\linewidth}
\vspace{-25pt}
\caption{Example data.}
\vspace{-4pt}
\begin{tabular}{@{}lllccc@{}}
\toprule
{\bf QuestionId} & {\bf UserId} & {\bf AnswerId} & {\bf AnswerValue} & {\bf CorrectAnswer} & {\bf IsCorrect} \\ \midrule
10322 & 452 & 8466 & 4 & 4 & 1 \\
2955 & 11235 & 1592 & 3 & 2 & 0 \\
3287 & 18545 & 1411 & 1 & 0 & 0 \\
10322 & 13898 & 6950 & 2 & 1 & 0 \\
\bottomrule
\end{tabular}
\label{tab:exampledata}
\end{table}

\begin{figure}
    \centering
    \includegraphics[width=0.7\linewidth]{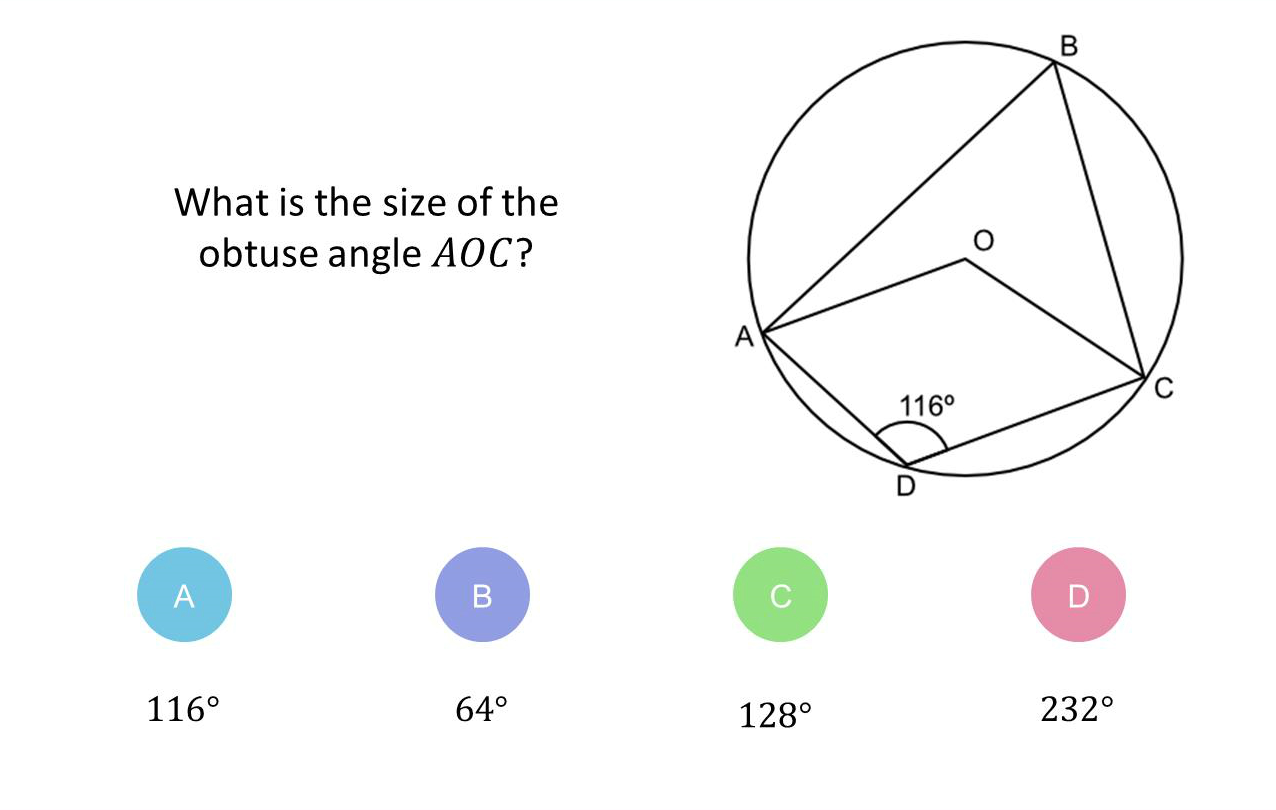}
    \caption{An example question from the education platform where the data we analyze is collected.}
    \label{fig:example_question}

\end{figure}

\noindent

We provide an extensive data provided by Eedi, an online education provider currently used in tens of thousands of schools, detailing student responses to multiple-choice diagnostic questions provided between September 2018 to May 2020. This platform offers crowd-sourced diagnostic questions to students from primary to high school (roughly between 7 and 18 years old). Each diagnostic question is a multiple-choice questions with 4 possible answer choices, exactly one of which is correct. Currently, the platform mainly focuses on mathematics questions. Figure~\ref{fig:example_question} shows an example question from the platform. All data is available to download from the competition homepage on CodaLab; see Section~\ref{sec:list}.

The competition is split into 4 tasks: tasks 1 and 2 share a dataset, as do tasks 3 and 4. These datasets are largely identical in format, but use disjoint sets of questions. All QuestionIds, UserIds and AnswerIds have been anonymized and have no discernable relation to those found in the product. Note that all such IDs for tasks 1 and 2 are anonymized separately from those for tasks 3 and 4: \textbf{IMPORTANT: Question, User and Answer IDs should not be linked between the data for these pairs of tasks!}. This is by design, to ensure that the two datasets are both self-contained.

\subsection{Primary Data}

This is main training data, consisting of records of answers given to questions by students. It can be found in the files \texttt{train\_task\_1\_2.csv} and \texttt{train\_task\_3\_4.csv} The columns are as follows:

\begin{itemize}
    \item \textbf{QuestionId:} ID of the question answered.
    \item \textbf{UserId:} ID of the student who answered the question.
    \item \textbf{AnswerId:} Unique identifier for the (QuestionId, UserId) pair, used to join with associated answer metadata (see below).
    \item \textbf{IsCorrect:} Binary indicator for whether the student's answer was correct (1 is correct, 0 is incorrect).
    \item \textbf{CorrectAnswer:} The correct answer to the multiple-choice question (value in [1,2,3,4]).
    \item \textbf{AnswerValue:} The student's answer to the multiple-choice question (value in [1,2,3,4]).
\end{itemize}

Table \ref{tab:exampledata} is an illustration of four data records in this format. Each student has typically answered only a tiny fraction of all possible questions and hence the matrix is extremely sparse. For tasks 1 and 2, we removed questions that have received fewer than 50 answers and students who have answered fewer than 50 questions. For tasks 3 and 4, where we are interested in a fixed set of questions, we removed all students who had answered fewer than 50 of these questions. In addition, when a student has multiple answer records to the same question, we keep the latest answer record. The data can be transformed into matrix form, where each row represents a student and each column represents a question.

For tasks 1 and 2, the individual answer records are randomly split into 80\%/10\%/10\% training/public test/private test sets. For tasks 3 and 4, the UserIds are randomly split into 80\%/10\%/10\% training/public test/private test sets. These preprocessing steps lead to training datasets of the following sizes:

\begin{itemize}
    \item Tasks 1 and 2: 27613 questions, 118971 students, 15867850 answers
    \item Tasks 3 and 4: 948 questions, 4918 students, 1382727 answers
\end{itemize}

The total number of answer records these training sets exceeds 17 million, rendering manual analysis impractical and necessitating a data-driven, machine learning approach. For an illustration of the matrix representation of the data, see Figures \ref{fig:task1data} and \ref{fig:task2data} in Section \ref{sec:tasks}.

\subsection{Question Metadata}

We provide the following metadata about each question:

\begin{itemize}
    \item \textbf{SubjectId} Each subject covers an area of mathematics, at varying degrees of granularity. We provide IDs for each topic associated with a question in a list. Example topics could include ``Algebra'', ``Data and Statistics'', and ``Geometry and Measure''. These subjects are arranged in a tree structure, so that for instance ``Factorising" is the parent subject of ``Factorising into a Single Bracket". We provide details of this tree in an additional file \texttt{subject\_metadata.csv} which contains the subject name and tree level associated with each SubjectId, in addition to the SubjectId of its parent subject.

    \item \textbf{Question content:}
    In Tasks 3 and 4, in addition to the topics, we will also provide the image presented to the student for each question, as shown in Figure \ref{fig:example_question}, for each of the questions. The question images have been shared solely for the purpose of this competition and must not be used for any other purpose. The question images must not be printed or shared with anyone outside of the competition. The question wording is contained in the images but will not be made available as text.
    
\end{itemize}

\subsection{Student Metadata}

The following metadata is provided about students in the dataset:

\begin{itemize}
    \item \textbf{UserId:} An ID uniquely identifying the student, which can be joined to the primary dataset.
    \item \textbf{Gender:} The student's gender, when available. 0 is unspecified, 1 is female, 2 is male and 3 is other.
    \item \textbf{DateOfBirth:} The student's date of birth, rounded to the first of the month.
    \item \textbf{PremiumPupil:} Whether the student is eligible for free school meals or pupil premium due to being financially disadvantaged.
\end{itemize}

\subsection{Answer Metadata}

The following metadata is provided about each individual answer record in the dataset:

\begin{itemize}
    \item \textbf{UserId:} An ID uniquely identifying the answer, which can be joined to the primary dataset.
    \item \textbf{DateAnswered:} Time and date that the question was answered, to the nearest minute.
    \item \textbf{Confidence:} Percentage confidence score given for the answer. 0 means a random guess, 100 means total confidence.
    \item \textbf{GroupId:} The class (group of students) in which the student was assigned the question.
    \item \textbf{QuizId:} The assigned quiz which contains the question the student answered.
    \item \textbf{SchemeOfWorkId:} The scheme of work in which the student was assigned the question.
\end{itemize}

\section{Task Details} \label{sec:tasks}

\noindent
In this section, we introduce the competition tasks and the evaluation metrics.

The competition consists of four tasks of varying styles.  The first two tasks aim to predict the student's responses to every question in the dataset. These two tasks can be formulated in several ways, for instance as a recommender system challenge \cite{harper2015movielens, bennett2007netflix, dror2012yahoo} or as a missing value imputation challenge \cite{little2019statistical, yoon2018gain, stekhoven2012missforest, gong2019icebreaker}. Here, each student only answers a small fraction of the questions while student responses to other questions are of great interest for personalized education. The third task focuses on evaluating question quality which is essential and remains an open question in the education domain \cite{wang2020large}. The final task directly addresses the challenge of personalized education where personalized dynamic decision making \cite{agarwal2016multiworld, li2010contextual, eddi} is needed.  
The first two tasks could serve as a useful basis on which to build solutions for the latter two tasks, but it is not necessary to take this approach. Participants are free to submit solutions to whichever tasks they wish and in whichever order they wish.

\begin{figure}[t]
    \centering

    \includegraphics[width=0.75\linewidth]{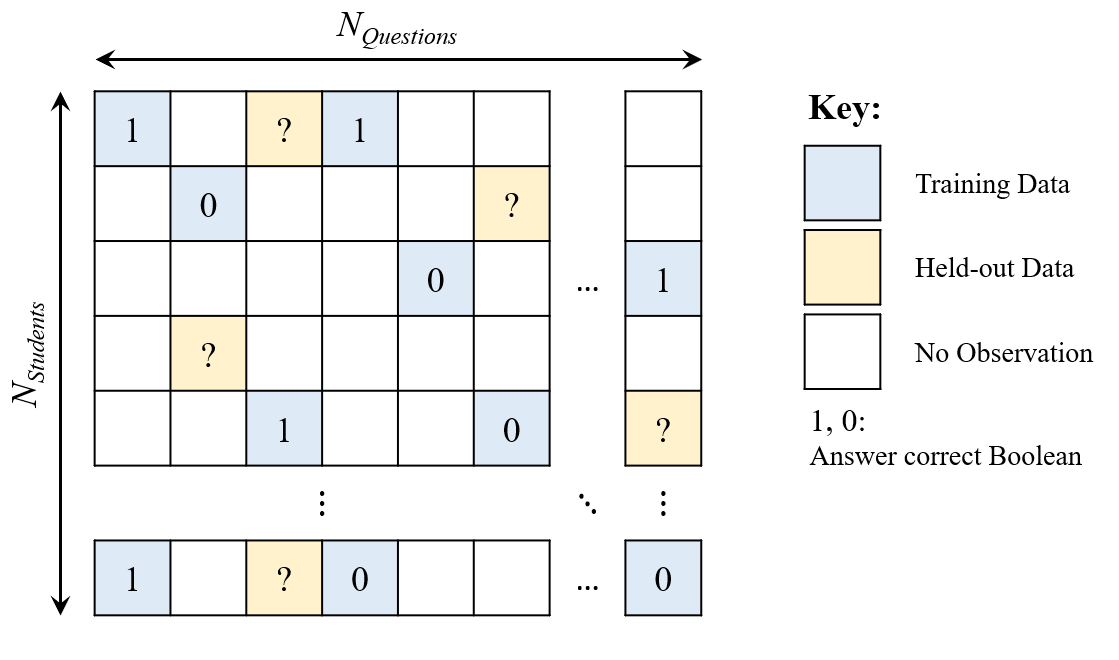}
    \caption{An illustration of the sparse matrix representation of the data for Task 1.}
    \label{fig:task1data}

\end{figure}

\subsection{Task 1: Predict Student Responses -- Right or Wrong} 
The first task is to predict whether a student answers a question correctly. The primary data used for this task is a table of records (\texttt{Student ID}, \texttt{Question ID}, \texttt{Is Correct} indicator) where the last column is binary valued. A sparse matrix representation of this data is illustrated in Figure \ref{fig:task1data}. The participants are asked to predict the correctness indicator on a held-out subset of the students' answers. More specifically, for each student, a portion of the available records will be held out as the hidden test set on which evaluation will be performed.

Predicting the correctness of a student's answers to not yet answered (or newly introduced) questions is crucial for estimating the student's ability level in a real-life personalized education platform, and forms the groundwork for more advanced tasks. This task falls under the class of matrix completion, and is reminiscent of challenges often seen in the recommender systems domain in the case of binary data. Popular approaches in this domain such as matrix factorization or nearest-neighbour based methods may prove effective at this task.

\subsubsection{Evaluation Metric}
We use {\bf prediction accuracy} as the metric, i.e., the number of predictions that match the true correctness indicator, divided by the total number of predictions (in the held-out test set): 
\begin{align*}
    {\rm Accuracy} = \frac{\#{\rm correct\,predictions}}{\#{\rm total\,predictions}}
\end{align*}

\subsection{Task 2: Predict Student Responses --  Answer Prediction:} 
\begin{figure}[t]
    \centering

    \includegraphics[width=0.75\linewidth]{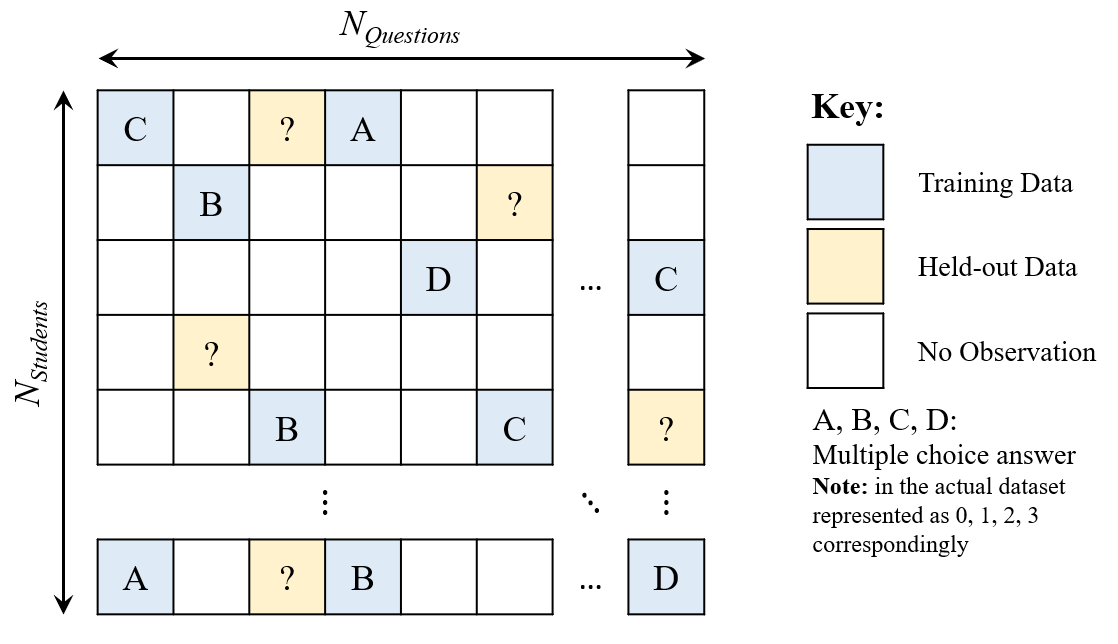}
    \caption{An illustration of the sparse matrix representation of the data for Task 2.}
    \label{fig:task2data}

\end{figure}

The second task is to predict which answer a student gave to a particular question. The primary data used for this task is a table of records (\texttt{StudentId}, \texttt{QuestionId}, \texttt{AnswerValue}, \texttt{CorrectAnswer}) where the last 2 columns are categorical taking values in [1, 2, 3, 4] (corresponding respectively to multiple-choice answer options A, B, C and D). The sparse matrix representation shown for Task 1 thus now looks as in Figure \ref{fig:task2data}. The questions in our dataset are all multiple-choice, each with 4 potential choices and 1 correct choice, so this task is a multi-class prediction problem -- the participants are asked to predict students' responses for a hidden, held-out subset of (StudentId, QuestionId) pairs.

Predicting the actual multiple-choice option for a student's answer allows analysing likely common misconceptions that a student may hold on a topic, and can thus form the basis for personalized advice and guidance on a real-life education platform. Clusters of question-answer pairs which are highly correlated may indicate that they correspond to the same, or related misconceptions. Understanding the relationships between misconceptions is a crucial problem to solve for curriculum development, it may inform the way a topic is taught and the sequencing of topics.

As in Task 1, this is a matrix completion task, but this time with unordered categorical data. Data of this type is rare in the recommender systems domain, where responses will typically be binary or ordinal (e.g. 1-5 stars), and so more novel approaches may be required in order to correctly predict students' answers and accurately model their misconceptions. The first two tasks form the foundation of analyzing students' learning, because most models that aim to produce such analytics rely on accurately modeling students and hence modelling their answers. Thus, participants competing in these two tasks are exposed to the fundamental task in educational data mining. 

\noindent

\subsubsection{Evaluation Metric}
We use the same metric {\bf prediction accuracy} as above, except that the true answers are now categorical instead of binary.

\subsection{Task 3: Global Question Quality Assessment} 

The third task is to predict the ``quality" of a question, as defined by a panel of domain experts (experienced teachers), based on the information learned from the students' answers found in the dataset. This task requires the definition of a metric for evaluating the question quality that mimics the experts' judgement of the question quality. Crucially, the experts' judgements will not be provided to the competition participants. The task is therefore very different in nature from the previous two, and is an unsupervised learning problem, demanding some innovative thinking. 

\begin{figure}[t]
    \centering
    \includegraphics[width=0.75\linewidth]{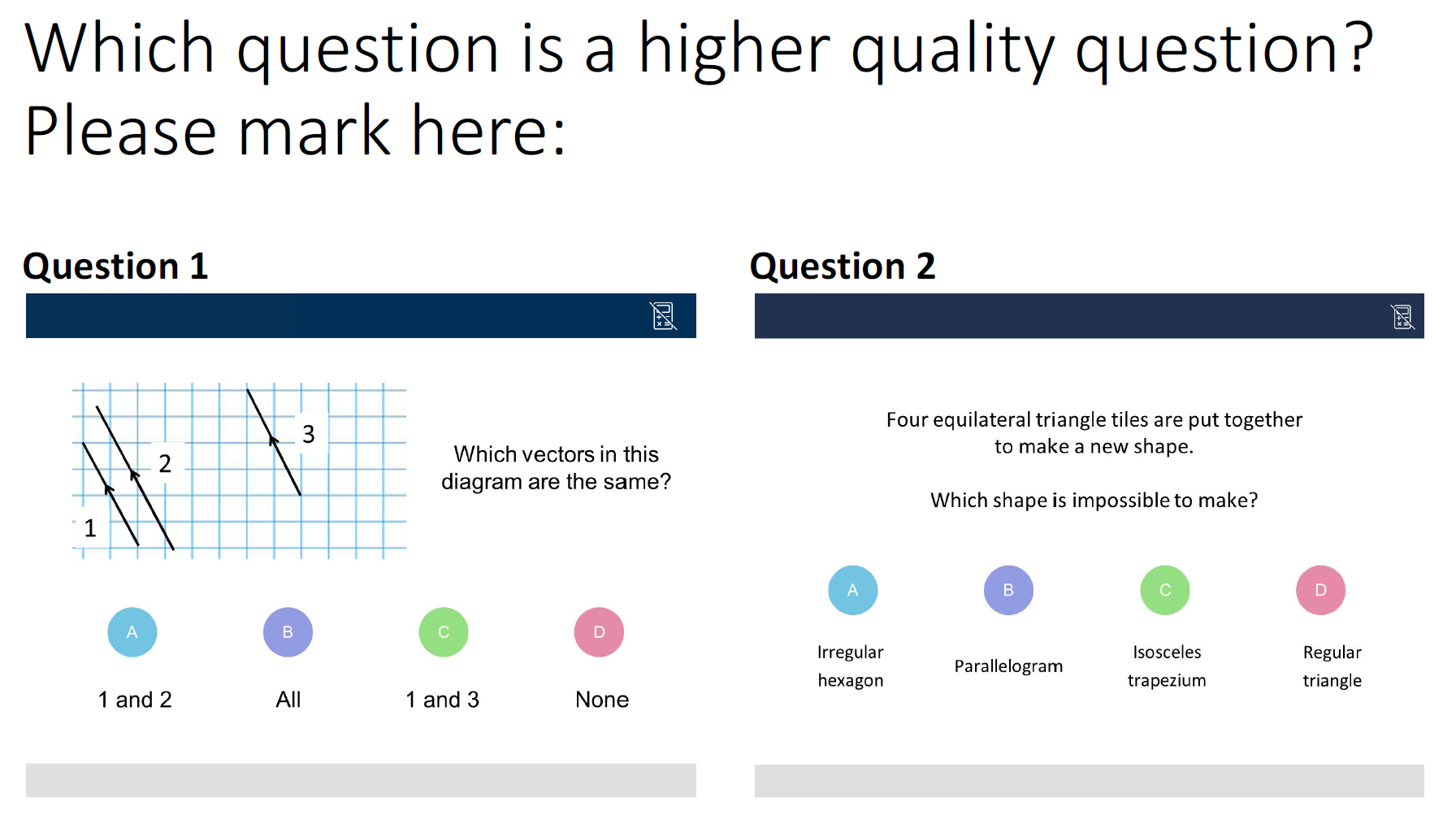}
    \caption{Example of a prompt used in collecting experts' judgement of pairwise relative question quality. In addition to this, the experts receive the following instructions: \textit{On each of the following slides, you will see 2 questions, one on the left and one on the right. Please decide which question is of higher quality; ties are not allowed.}}
    \label{fig:task3expert}
\end{figure}

In order to guide the participants, we present a note on how expert teachers judge question quality, including their intuition and the criteria they use. The participants may use this material to design their automatic question quality metrics. An example of a prompt used in the expert data collection is shown in Figure \ref{fig:task3expert}. In addition, the following ``Golden rules" of quality question design have been identified by one of the domain experts, Craig Barton\footnote{\url{https://medium.com/eedi/what-makes-a-good-diagnostic-question-b760a65e0320}}: 
\begin{itemize}
    \item They should be clear and unambiguous
    \item They should test a single skill/concept
    \item Students should be able to answer them in less than 10 seconds
    \item You should learn something from each incorrect response without the student needing to explain
    \item It is not possible to answer the question correctly whilst still holding a key misconception.
\end{itemize}

The question quality metric designed in this task is of paramount importance in crowd-sourced education applications, as it provides a scalable way to evaluate the quality of the questions submitted by teachers. The quality of crowd-sourced questions reflects directly on the usefulness of the platform to the students and teachers. The quality judgement can also be used for personalized guidance for the teachers, helping them to improve question quality.

\begin{figure}
    \centering
    \includegraphics[width=1.0\linewidth]{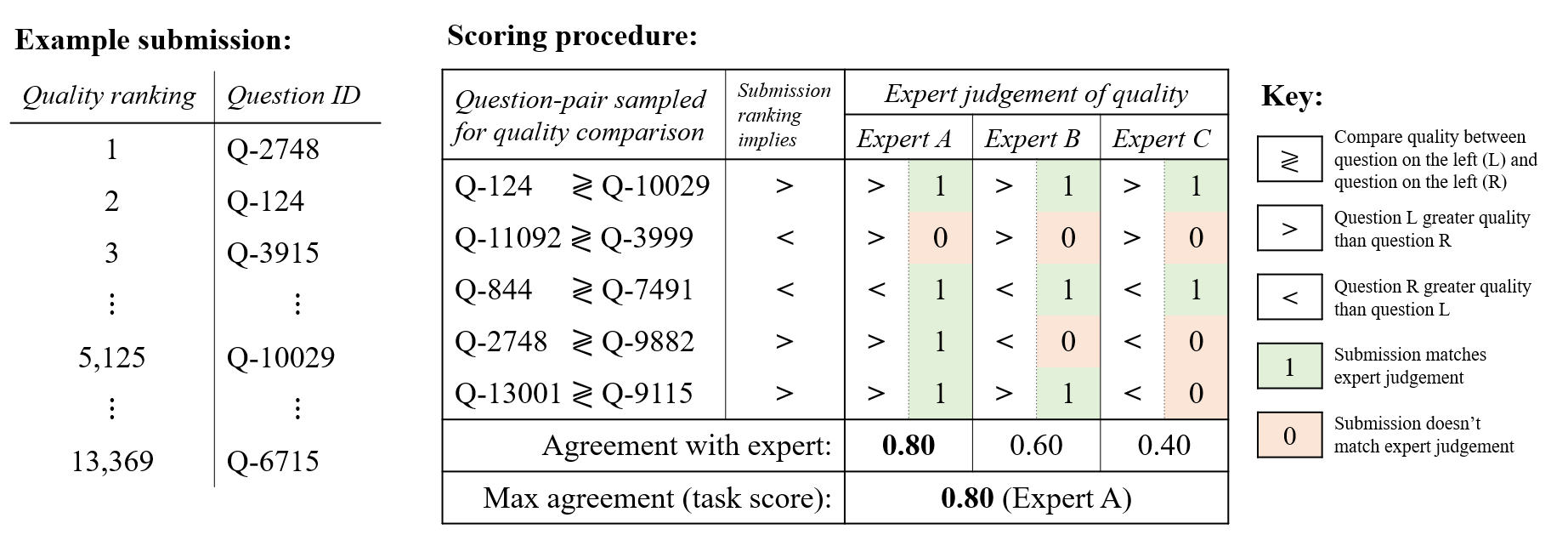}
    \caption{An illustration of the scoring process in Task 3. \textit{Left}: The expected format of the submissions for Task 3 - a ranking of question quality over the \texttt{Question ID}s, in the decreasing order of quality. \textit{Right}: An illustration of the performance metric calculation, see the steps in the main body text (Section \ref{task3eval}). This example uses 5 question-pairs and 3 experts.}
    \label{fig:task3metric}
\end{figure}

The competitors are encouraged to utilise the machine learning model(s) used in the previous task(s) in defining this metric. Based on this metric, the participants must provide a ranking: rank from $1$ to $N_{Questions}$, each uniquely mapping to a \texttt{Question ID} in the dataset. Rank $1$ should correspond to the highest quality question, and so on, in order of decreasing question quality. The absolute values of the metric are not required. For an illustration of the required output see Figure \ref{fig:task3metric}. The evaluation procedure and metric are described in Section 1.5. sw{There is no 1.5}

This task can be viewed as an unsupervised learning task, as there is no explicit supervision label available for question quality. Insights from areas such as information theory, feature selection and learning to rank may be relevant to this task.

\subsubsection{Evaluation Metric} \label{task3eval}
The participants will submit a quality ranking for the questions (Figure \ref{fig:task3metric} \textit{Left}). An unseen set of question-pairs will then be used to evaluate the quality of this ranking (and thus the underlying metric). Note that we have collected the experts' judgement of which question in each pair is of higher quality. The evaluation steps are then as follows (see also Figure \ref{fig:task3metric} \textit{Right} for an illustration):
    \begin{itemize}
        \item Determine, based on the submitted ranking, which question in each pair is of higher quality.
        \item Compare this with each expert's judgement (assign 1 if matching, 0 if not matching).
        \item For each expert $i$, determine the agreement fraction: $A_i = \frac{N_{matching\_pairs}}{N_{total\_pairs}}$ .
        \item Find the \textit{maximum} of these agreement fractions $A_{max} = \max_{i} A_i$. This will be used as the final evaluation metric for this task.
    \end{itemize}
    
    We are looking for metrics that can approximate \textit{an} expert judgement really well, hence we use the maximum of the agreement fractions, rather than a mean of the agreement fractions over all experts. The reasoning for this approach is that the quality metrics of the experts are in themselves subjective, and it is interesting to find whether a particular expert's approach can be approximated especially well by the use of machine learning.

\subsection{Task 4: Personalized Questions} The fourth task is to interactively generate a sequence of questions to ask a student in order to maximise the predictive accuracy of a model on their remaining answers. Specifically, a participant's model will be provided with a set of previously-unseen students, whose answers to questions are completely hidden, and a set of potential questions to query for each student. The model will then choose a personalized question to query for each of these students in turn, and then their corresponding answer will be revealed to the model. Based on this information, the model should choose a second question to query for each student, and so on, until 10 questions have been asked in total. 

The aim of the task is to maximise the predictive accuracy of a participant's model on a held-out set of questions for each student, after the model has been exposed to the 10 answers from each student. This task is of fundamental importance to personalized education, where we wish to accurately diagnose a student's level of understanding of various concepts while asking the minimum number of questions possible, in order to make the most efficient use of both student and teacher time. The task is also a crucial machine learning challenge, requiring participants to reason effectively about their model's uncertainty, and to use data as efficiently as possible.

This task can be viewed through the lens of a number of related fields, including active learning, reinforcement learning, bandit algorithms, Bayesian experimental design and Bayesian optimization, and insights drawn from any of these fields will likely prove useful.

\begin{figure}[t]
    \centering
    \includegraphics[width=0.75\linewidth]{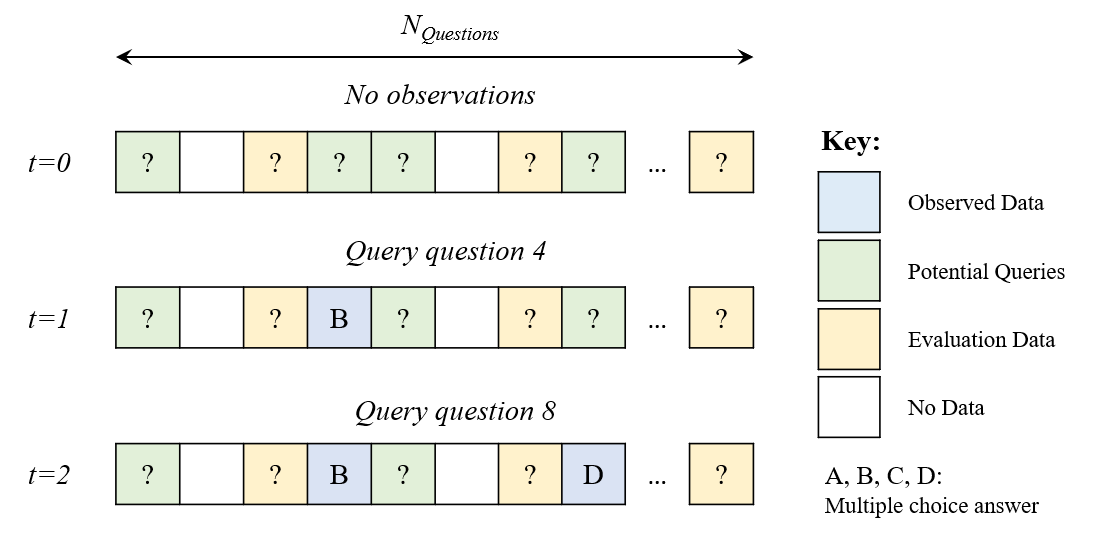}
    \caption{An illustration of the procedure for Task 4. On each time step, the model is able to train on the data in blue, and its predictive performance is assessed on the held-out data in yellow. The algorithm must then choose the next question to query from the set of green questions using this new model.
    }
    \label{fig:task4data}
\end{figure}

\subsubsection{Evaluation Metric}
Submitted models will be asked to sequentially choose 10 query questions for every student in a held-out set of students. After each selection step,  both the categorical answer and binary correctness indicator for these student-question pairs will be revealed to the model in private. The model is then given the opportunity to incorporate this new data or retrain after each question. After receiving 10 answers for each student, the model will be assessed on its prediction accuracy for predicting the binary correctness indicators for a held-out test set of answers for each of these students, that cannot be queried. \

\section{Submission Protocol}
\label{sec:submission}

Each task contains two phases: a public evaluation phase and a private evaluation phase. Results in the public evaluation phase are displayed on a public leaderboard allowing participants to see how their submissions perform compared to other participants. Results in the private evaluation phase are hidden until the end of the competition. \textbf{Important: For each task, participants must submit to both the public and private evaluation phases separately. Submissions made solely to the public evaluation phase will not be used in the final judging of the competition. It is the participants' responsibility to make sure their submission to the private phase of each task represent their best results.} For tasks 1-3, submission template files are provided for both the public and private phases, which must be submitted to CodaLab. For task 4, a participants model must be submitted to both the public and private leaderboards, where it will be evaluated separately.

Before submission, the participants must first zip their submission files into one zip file. CodaLab only accepts one single \texttt{.zip} file as the submission file. 
The submission buttons are under \texttt{Participate} tab on the top of the competition website and under the \texttt{Submit/View Results} tab on the left. The tabs on the top of this page, e.g., \texttt{Task 1 Public}, contains the submission link to each of the tasks.

Below are the detailed submission instructions for each task.

\subsection{Submission for Tasks 1-3}
The first three tasks are evaluated through the submission of a zipped CSV file to Codalab:
\noindent
\begin{itemize}
    \item \textbf{Task 1}: Participants are provided with a CSV file (in the folder \texttt{starter\_kit/submission\_templates}) containing (UserId, QuestionId) pairs for which the answer is unseen. They must fill in their prediction for whether the student will answer the question correctly in each case. 
    {\bf Important: the submission file must be a \texttt{.zip} file containing a file named \texttt{submission\_task\_1.csv}}.
    \item \textbf{Task 2}: Participants are provided with a CSV file (in the folder \texttt{starter\_kit/submission\_templates}) containing (UserId, QuestionId) pairs for which the answer is unseen. They must fill in their prediction for which answer the student will give to the question in each case. 
    {\bf Important: the submission file must be a \texttt{.zip} file containing a file named \texttt{submission\_task\_2.csv}}.
    \item \textbf{Task 3}: Participants are provided with a CSV file (in the folder \texttt{starter\_kit/submission\_templates}) containing a list of all of the QuestionIds provided in the dataset for Task 3. They are asked to fill in a 'Ranking' column indicating the rank (1-948) they give to each question, where 1 is the highest-quality question and 948 is the lowest-quality question.
    {\bf Important: the submission file must be a \texttt{.zip} file containing a file named \texttt{submission\_task\_3.csv}}.
\end{itemize}
\subsection{Submission for Task 4}\label{task_4_submission}
Task 4 is evaluated through a code submission. Participants are provided with a template file \texttt{submission\_model\_task\_4.py}, which provides a simple API wrapper through which the evaluation script will interface with submitted models. The methods which participant must implement are as follows:
\begin{itemize}
    \item \textbf{\_\_init\_\_}: Load a participant's model from this file or a neighbouring file (e.g. a separate \texttt{model.py} file where the model is defined) and perform any initialisation that is required.
    \item \textbf{select\_questions}: Select the next question to query for each question, given the data observed by the model so far (both the binary correctness indicators and the specific multiple-choice answers) and an array indicating which questions can be queried for each student.
    \item \textbf{update\_model}: Optionally update the model based on the new data revealed after revealing new answers.
    \item \textbf{predict}: Predict a binary correctness indicator for each student, for all questions for which we have not observed the student's answer.
\end{itemize}

Details of the specific signatures of these functions can be found in the template \texttt{submission\_model\_task\_4.py} file. The evaluation procedure is as follows:
\begin{enumerate}
    \item Initialise the model with \texttt{\_\_init\_\_()}.
\item For 10 steps, first select a new feature with \texttt{select\_questions()}, reveal the selected values, and pass the new data to \texttt{update\_model()}.
\item After 10 feature selection steps, call \texttt{predict()} to make binary predictions for the held-out target elements and evaluate the model's prediction accuracy.
\end{enumerate}

The submission files for this task must contain \texttt{submission\_model\_task\_4.py}.
Users are free to include additional files, such as model definitions or trained model weights, in their submissions.
To submit, the \texttt{submission\_model\_task\_4.py} template file and any additional files should be zipped into one single \texttt{.zip} file and then submitted to CodaLab. 
{\bf Important: please \textit{do not} change the class method names in \texttt{submission\_model\_task\_4.py}. Also, for any saved model files, participants \textit{must} use \texttt{model\_task\_4\_} as the prefix of the file name for submission}. For example, if the model is called \texttt{my\_pytorch.pt}, then for submission , this model file must be renamed to \texttt{model\_task\_4\_my\_pytorch.pt}.

During evaluation, all training data and metadata files included as part of task 4 will be added to the root of the submission directory, 
and submissions may make use of these files as they wish. Simply specifying the name of the dataset to load should be sufficient to load the files -- for instance, the training data can be accessed by submitted models at the path ``\texttt{./train\_task\_3\_4.csv}''. Note that users are expected to upload a trained model for submission, as time limits on the submission compute workers will likely render training prior to evaluation too time-consuming.

For each task, there is a daily and total submission limit specified on the relevant submission page on Codalab.  Unsuccessful submissions due to errors will not count against this total. \textbf{Submissions must be made separately to both the public and private components of the leaderboard}, with the final competition results being based solely on the private leaderboard.

\subsection{Leaderboard}
The submitted result will show up in a public leaderboard for each public phase of the competition. The public leaderboard shows how the participant(s) stand against other participants on the public evaluation data for each task. 

Note that the leaderboard only shows the same ``score'' column for all tasks; participants should be aware that the meaning of the ``score'' is different for each task; please refer to the details of the evaluation metrics for each task in Section~\ref{sec:tasks}. Nevertheless, a more detailed result can be accessed in the ``detailed results'' part on the rightmost column of the table under the ``Results'' tab of the competition website.

For all private phases of the competition, no public leaderboard will be shown and results are only visible by the competition organizers.

\subsection{Computation Environment}
The evaluation on CodaLab is performed using the an off-the-shelf docker image that contains most of the data science, machine learning and deep learning packages. Participants must ensure that their submissions are able to run in this environment. Please see \url{https://github.com/ufoym/deepo} for more detail.
\section{Getting Started: Sample Model, Local Evaluation and Submission Preparation}
\label{sec:get-started}

\subsection{Quick Start}

Both the public data and starter kit for the competition can be found under the tab \texttt{Participate}/\texttt{Get Data}/ from the competition homepage. The starter kit contains a number of utility scripts and sample models to allow easy participation in the competition. Submission templates for preparing submissions in the correct format for each task are included in the \texttt{submission\_templates} directory.

To quickly get started with the competition, follow these instructions:
\begin{enumerate}
   \item Download the training data and starter kit from the competition homepage.
   \item Place the downloaded \texttt{data} directory into the root of the \texttt{starter\_kit} directory.
   \item (Optional) Run the \texttt{local\_data\_split.py} files for each task in order to generate validation sets for local model evaluation.
   \item Run the sample model files provided for tasks 1-3 in order to generate sample submissions for the competition. Task 4 requires no generation, as participants must submit the model code itself.
\end{enumerate}

To submit solutions to Tasks 1-3, participants should submit a \texttt{.zip} file containing a completed submission template file named \texttt{submission\_task\_n.csv} where \texttt{n} is the task number. This file should then be uploaded to \textbf{both} the public and private phases of the appropriate task.

To submit solutions to Task 4, participants should submit a \texttt{.zip} file containing a completed submission API wrapper file \texttt{submission\_model\_task\_4.py}, in additional to any additional model files or artifacts required in order to run the trained model.

Each task has its own self-contained directory, which typically includes a script for creating a local ``validation set'' for local model evaluation, a script for performing local model evaluation, and an example model to help get started with the task.

Further details for each task's resources are provided in the following sections.

\subsection{Task 1}
\paragraph{Available Files} The following files are available for Task 1:
\begin{itemize}
    \item Training data: \texttt{train\_task\_1\_2.csv}, in the folder \texttt{data/train\_data}
    \item Submission template: \texttt{submission\_task\_1\_2.csv}, in the folder \\ \texttt{starter\_kit/submission\_templates}
     \item Question metadata: \texttt{question\_metadata\_task\_1\_2.csv}, in the folder  \texttt{data/metadata}
    \item Student metadata: \texttt{student\_metadata\_task\_1\_2.csv}, in the folder  \texttt{data/metadata}
    \item Answer metadata: \texttt{answer\_metadata\_task\_1\_2.csv}, in the folder  \texttt{data/metadata}
    \item Local evaluation scripts: \texttt{local\_data\_split.py}, \texttt{local\_evaluation.py}, in the folder\\ \texttt{starter\_kit/task\_1}
    \item Sample baseline model: \texttt{sample\_model\_majority.py}, in the folder \texttt{starter\_kit/task\_1}
\end{itemize}

\paragraph{Submission to CodaLab}
To submit to CodaLab, participants must submit results containing predictions for {\it each and every} \texttt{(UserId, QuestionId)} pair in the submission template file \texttt{submission\_task\_1\_2.csv}. Running the provided \texttt{sample\_model\_majority.py} script will generate an example of this file, creating a directory \texttt{../submissions} by default where the generated file is named as \texttt{submission\_task\_1.csv} which is ready to be zipped and submitted.

{\bf Important}: make sure the name of the prediction column in submissions is \texttt{IsCorrect}. 

\paragraph{Local Evaluation} 

To evaluate locally on a validation set produced from the training data, follow the steps below:
\begin{enumerate}
    \item Navigate into the folder \texttt{starter\_kit/task\_1}
    \item Split the data. We have provided a script \texttt{local\_data\_split.py} to do so; participants can perform data split by running \\
    \centerline{\texttt{python local\_data\_split.py}}\\ which generates a train and validation set using the training data. By default, the split files are named as \texttt{train\_task\_1\_2.csv} and \texttt{valid\_task\_1\_2.csv} and are saved in the folder \texttt{data/test\_input}.
    \item Run your model and make predictions. We have provided a sample model \\ \texttt{sample\_model\_majority.py} to help get started. Note that in order to run this model on the local evaluation data split, the block of code marked ``Default arguments for Codalab submission" should be commented out, and the block marked ``Default arguments for local evaluation" should be uncommented. To get predictions using this model, please run \\
    \centerline{\texttt{python sample\_model\_majority.py}} \\ This model outputs a \texttt{.csv} file containing the results at \\(\texttt{data/test\_input/test\_submission\_task\_1.csv}).
    \item Evaluation. Run \\
    \centerline{\texttt{python local\_evaluation.py} }\\with appropriate options; see \texttt{argparse} arguments. This command computes and saves the score using the prediction and validation set. A score and confusion matrix are saved to the  \texttt{output\_dir} which by default is \texttt{data/test\_output}. The official evaluation for this task implements the same evaluation metric as that in \texttt{local\_evaluation.py}.
\end{enumerate}

\subsection{Task 2} 
\paragraph{Available Files}
The following files are available for Task 2:
\begin{itemize}
    \item Training data: \texttt{train\_task\_1\_2.csv}, in the folder \texttt{data/train\_data}
    \item Submission template: \texttt{submission\_task\_1\_2.csv}, in the folder\\ \texttt{starter\_kit/submission\_templates}
    \item Question metadata: \texttt{question\_metadata\_task\_1\_2.csv}, in the folder  \texttt{data/metadata}
    \item Student metadata: \texttt{student\_metadata\_task\_1\_2.csv}, in the folder  \texttt{data/metadata}
    \item Answer metadata: \texttt{answer\_metadata\_task\_1\_2.csv}, in the folder  \texttt{data/metadata}
    \item Subject metadata: \texttt{subject\_metadata.csv}, in the folder  \texttt{data/metadata}
    \item Local evaluation scripts: \texttt{local\_data\_split.py}, \texttt{local\_evaluation.py}, in the folder \\\texttt{starter\_kit/task\_2}
    \item Sample baseline model:  \texttt{sample\_model\_majority.py}, in the folder \texttt{task\_2}
\end{itemize}
\paragraph{Local Evaluation and Prepare Submission}
Since Task 2 is very similar in nature to Task 1, the training data and submission templates are the same and the local evaluation scripts are similar. The only difference is that Task 2 predicts the actual response that a student makes to a question (answer A, B, C or D to each multiple-choice question, encoded as 1, 2, 3 or 4 respectively).

{\bf Important}: make sure the name of the prediction column in the submission file is \texttt{AnswerValue}.

\subsection{Task 3}
\paragraph{Available Files}
The following files are available for Task 3: 
\begin{itemize}
    \item Training data: \texttt{train\_task\_3\_4.csv}, in the folder \texttt{data/train\_data}
    \item Submission template: \texttt{submission\_task\_3.csv}, in the folder \\ \texttt{starter\_kit/submission\_templates}
    \item Question images: in the folder \texttt{data/images/}
    \item Question metadata: \texttt{question\_metadata\_task\_3\_4.csv}, in the folder  \texttt{data/metadata}
    \item Student metadata: \texttt{student\_metadata\_task\_3\_4.csv}, in the folder  \texttt{data/metadata}
    \item Answer metadata: \texttt{answer\_metadata\_task\_3\_4.csv}, in the folder  \texttt{data/metadata}
    \item Subject metadata: \texttt{subject\_metadata.csv}, in the folder  \texttt{data/metadata}
    \item Sample baseline model: \texttt{sample\_model\_entropy.py}, in the folder \texttt{starter\_kit/task\_3}

\end{itemize}

\paragraph{Submission to CodaLab}
This task asks participants to rank the quality of questions in the training data \texttt{train\_task\_3\_4.csv} in descending order, i.e., rank 1 represents the highest quality, rank 2 represents the second highest quality, etc. There is no ground-truth provided and local evaluation is not possible; rather, we provide a baseline model which ranks the questions based on an estimation of their entropy (\texttt{sample\_model\_entropy.py}) to generate results appropriate for submission for Task 3. See the implementation in this files for details of the computations. The submission file contains 2 columns \texttt{(QuestionId, ranking)} where the second column is the ranking for each question. Each question should have a unique ranking, i.e., no two rankings should be the same. 

The training data and metadata provided for tasks 3 and 4 is in the same format as that of tasks 1 and 2, but uses a disjoint, smaller set of questions. As explained in section \ref{sec:data}, the randomized IDs used in Tasks 1 and 2 are generated independently of those used in Tasks 3 and 4, and so participants should not attempt to use the data from Tasks 1 and 2 to aid them in Tasks 3 and 4.

To prepare submission run your model which generates the ranking for all question IDs in the \texttt{submission\_task\_3.csv} file. The provided entropy-based model demonstrates this process; simply run\\
 \centerline{\texttt{python sample\_model\_random.py}}\\
This will create a prediction file \texttt{test\_submission\_task\_3.csv} in the \texttt{../submissions} folder by default. This file is ready to be zipped and submitted.

{\bf Important}: the ranking column must be named \texttt{ranking} for the CodaLab evaluation script to correct read the participants' predicted quality rankings.

\subsection{Task 4}
\paragraph{Available Files}
The following files are available for Task 4:
\begin{itemize}
    \item Training data: \texttt{train\_task\_3\_4.csv}, in the folder \texttt{data/train\_data}
    \item Submission template file: \texttt{submission\_task\_4.py}, in the folder \texttt{submission\_templates}
    \item Question images: in the folder \texttt{data/images/}
    \item Question metadata: \texttt{question\_metadata\_task\_3\_4.csv}, in the folder  \texttt{data/metadata}
    \item Student metadata: \texttt{student\_metadata\_task\_3\_4.csv}, in the folder  \texttt{data/metadata}
    \item Answer metadata: \texttt{answer\_metadata\_task\_3\_4.csv}, in the folder  \texttt{data/metadata}
    \item Subject metadata: \texttt{subject\_metadata.csv}, in the folder  \texttt{data/metadata}
    \item Local evaluation scripts and models: 
    \vspace{-5pt}
    \begin{itemize}
        \item \texttt{local\_data\_split.py} is in the folder \texttt{starter\_kit/task\_4} 
        \item \texttt{local\_evaluation.py}, \texttt{train\_model.py}, \texttt{submission\_model\_task\_4.py},\\ \texttt{model\_task\_4.pt}, \texttt{model.py} in the folder \texttt{starter\_kit/task\_4/pytorch}
         \item \texttt{local\_evaluation.py}, \texttt{train\_model.py}, \texttt{submission\_model\_task\_4.py},\\ \texttt{model\_task\_4\_most\_popular.npy}, \texttt{model\_task\_4\_num\_answers.npy}\\
         \texttt{model.py}, in the folder \texttt{starter\_kit/task\_4/numpy}
    \end{itemize}

\end{itemize}

\paragraph{Submission to CodaLab}
As detailed in \nameref{task_4_submission}, participants should submit to CodaLab a \texttt{.zip} file containing their completed \texttt{submission\_model\_task\_4.py} file along with any supplementary model code or weights, where it will be evaluated in private. Submissions may assume that all training data and metadata files for the task will be included same directory as the submission files (i.e., \texttt{./}).

We have provided sample models and script to generate a sample submission ready for CodaLab. To do so, please follow the steps below:
\begin{enumerate}
    \item Navigate to either \texttt{starter\_kit/task\_4/numpy} or \texttt{starter\_kit/task\_4/pytorch}
    \item Run \\\centerline{\texttt{python train\_model.py}}\\
    which saves a model file to either the NumPy or PyTorch folder.
    \item If using the NumPy model, zip \texttt{model.py}, \texttt{submission\_model\_task\_4.py}, \texttt{model\_task\_4\_\\most\_popular.npy} and \texttt{model\_task\_4\_num\_answers.npy} to a \texttt{.zip} file. If using the PyTorch model, zip \texttt{model.py}, \texttt{submission\_model\_task\_4.py} and \texttt{model\_task\_4.pt} into a \texttt{.zip} file. Details on the naming of the files and what can be changed inside the submission template file \texttt{submission\_model\_task\_4.py} are discussed in Section~\ref{task_4_submission}.
\end{enumerate}

\paragraph{Local Evaluation}
In the starter kit for Task 4, a mock test environment is provided as part of the competition API, in order to allow the entrants to check that their submission will work with the evaluation procedure. Both the NumPy and PyTorch models described above can be evaluated locally in this way.
\begin{enumerate}
    \item Navigate into the folder \texttt{starter\_kit/task\_4}
    \item Split the data. We have provide a script \texttt{local\_data\_split.py} to do so; participants can perform data split by running \\
    \centerline{\texttt{python local\_data\_split.py}}\\ which generates a train and validation set using the training data. By default, the split files are named as \texttt{train\_task\_4.csv} and \texttt{valid\_task\_4.csv} and are saved in the folder \texttt{data/test\_input}.
    \item Train your model. We have provided a sample model file and training scripts using both NumPy and PyTorch. To train the sample models, navigate to either \texttt{starter\_kit/\\task\_4/numpy} or \texttt{starter\_kit/task\_4/pytorch} and run \\ \centerline{\texttt{python train\_model.py}}\\
    Note that in the NumPy model, the path to the training data in the \texttt{train\_model()} method will need to updated to point to the newly created ``local test" split.
    This script will save a model file to the NumPy or PyTorch folders above; see the \texttt{train\_model.py} script for more details. Note that the provided PyTorch model class does not implement a method to train the model; 
    \item Evaluate. Run \\
    \centerline{\texttt{python local\_evaluation.py} }\\ This command computes and prints the sequence of selected questions and the final accuracy.
\end{enumerate}

\bibliographystyle{plain}
\bibliography{ref.bib}

\end{document}